\documentclass{pasj00}
\draft

\begin{document}
\SetRunningHead{Author(s) in page-head}{Running Head}

\title{The effect of the broad-line region with geometrical structures on $\gamma$-ray absorption in blazars}


\author{Maichang Lei}
\affil{Yunnan Observatory, Chinese Academy of Sciences,  Kunming 650011, China}
\email{maichanglei83@ynao.ac.cn}

\author{Jiancheng Wang}
\affil{Yunnan Observatory, Chinese Academy of Sciences,  Kunming 650011, China}
\email{jcwang@ynao.ac.cn}

%


\maketitle

\begin{abstract}
The broad-line region\,(BLR) is an important component of blazars, especially for the flat spectrum radio quasars\,(FSRQs). The soft photons arising from the BLR will substantially affect the transparency of the $\gamma$-ray photons produced in the relativistic jet. In the paper, we study the effect of the geometrical structure of the BLR on the absorption of $\gamma$-rays. We find that the $\gamma$-ray optical depth strongly depends on the geometrical structure of the BLR. For a ``flat" geometry of the BLR, the $\gamma$-ray photons with specified energies could escape transparently even their emission region locates inside the cavity of the BLR.
\end{abstract}

\section{Introduction}
Blazars are generally considered as the most extreme class of radio-loud active galactic nuclei\,(AGNs), in which the two-sided jets with magnetized plasmas are launched nearby the central engine at relativistic velocity aligned with the line of sight. Traditionally, blazars are classified as BL Lacs and FSRQs according to the equivalent width\,(EW) of their strong emission lines \citep{1995PASP..107..803U}. The objects with a rest frame $ EW<5 \AA$ are called BL Lacs, otherwise called FSRQs. New, based on the location of synchrotron peak frequency\,($\nu_{\rm peak}^{\rm S}$) \citep{2010ApJ...716...30A}, blazars with values of $\nu_{\rm peak}^{\rm S}\lesssim10^{14}$\,Hz are named low synchrotron peaked blazars\,(LSP), ones with values between $10^{14}$\,Hz$\lesssim\nu_{\rm peak}^{\rm S}\lesssim10^{15}$\,Hz are named intermediate synchrotron peaked blazars\,(ISP), others with values of $10^{15}$\,Hz$\lesssim\nu_{\rm peak}^{\rm S}$ are named high synchrotron peaked blazars\,(HSP). Otherwise, from the ratio of luminosity of the BLR in units of Eddington units, e.g., $L_{\rm BLR}/L_{\rm Edd}$, blazars with $L_{\rm BLR}/L_{\rm Edd}>10^{-3}$ mainly display the properties of FSRQs \citep{2011MNRAS.414.2674G}. Actually, these classifications are potentially associated with the BLR.

The existence of the BLR around central engine is an almost uncontroverted fact in blazars, especially for FSRQs. The BLR is known to emit a number of strong lines and the diffuse continuums at the optical to UV bands, but its structure, for example, with spherical or ``flat" geometry, is still under debate. Usually a spherical geometry of the BLR is assumed, in which its photons follow a blackbody spectrum peaking at the frequency of $\nu_{{\rm Ly}{\alpha}}=2.47\times 10^{15}$\,Hz \citep{2007MNRAS.375..417C,2008MNRAS.386..945T,2009MNRAS.397..985G}, and its radius $R_{\rm BLR}$ is determined by the reverberation mapping technique \citep{1982ApJ...255..419B} derived from its reprocessing emission of the accretion disk \citep{2000ApJ...533..631K,2005ApJ...629...61K,2007ApJ...659..997K,2009ApJ...697..160B,2008MNRAS.387.1669G}. If the $\gamma$-ray emission region of blazars is located within the cavity of the BLR, the photons from the BLR will significantly absorb the $\gamma$-rays with energies above few tens of GeV via positron-electron pair production processes \citep{2003APh....18..377D,2009MNRAS.392L..40T,2012arXiv1209.2291T}. \citet{2006ApJ...653.1089L} used a spherical BLR with half thickness $h$\,(0.05$\to 0.3$\,pc) to investigate the absorption of $\gamma$-rays, and demonstrated that the $\gamma$-rays with energies between 10 and 200\,GeV cannot escape from the diffuse photon field of the BLR.

For the geometry of the BLR, \citet{2006MNRAS.373..551L} found that the spherical BLR fails to explain the observed line width and shape in quasi-stellar objects. The ``black hole mass deficit" observed in NLS1s \citep{2004ApJ...606L..41G,2000NewAR..44..419H} disappears provided that NLS1s have a disc-like, rather than spherical BLR \citep{2008MNRAS.386L..15D}. \citet{2009MNRAS.392L..40T} produced a reasonable hardening of spectrum of quasar 3C 279 in TeV bands assuming anisotropic BLR structure. The stability of GeV breaks, as well as the $\gamma$-ray spectral breaks from a fraction of GeV to tens of GeV band in blazars can be well produced by the absorption of He II Lyman recombination continuum and emission lines, indicating that $\gamma$-rays arise within the high-ionization zone of the BLR \citep{2010ApJ...717L.118P}. \citet{2011MNRAS.417L..11S} studied the spectral properties of 3C 454.3 in GeV bands, implying that $\gamma$-ray emission region is close to the inner boundary of the BLR, which is impossible for the spherical BLR. Recently, \citet{2011MNRAS.413...39D} studied the properties of the BLR for a set of blazars and reference quasars, and supported that the BLR could have a flat geometry in blazars.

In the paper, we study the effect of the geometrical structure of the BLR on the absorption of $\gamma$-rays in detail. Section 2, we characterize our model. Subsection 2.1, the geometry and emission of the BLR are presented, and the $\gamma$-ray optical depth is calculated in Subsection 2.2. Our results are presented in Section 3. Discussion and conclusions are given in Section 4.

\section{The Model}
\subsection{Geometry and the BLR emission}
In the following, we assume that the $\gamma$-ray emission region moves at relativistic velocity along the jet and  locates at a position $R_{\rm o}$ above the accretion disk at a certain time. The generalized geometry of the BLR is shown in Figure \ref{fig:BLR}. Two shaded areas with axis symmetry represent the BLR clouds, and the central strip corresponds to the accretion disk. The BLR is assumed to be flat and characterized by the aperture angle $\alpha$ measured from the disk plane. In calculations, $\alpha$ are set between $15^{\circ}$ and $85^{\circ}$. We only consider the photons from the BLR without being obscured by accretion disk. The BLR is geometrically thick with the scale from inner radius $r_{\rm BLR,i}$ to outer radius $r_{\rm BLR,o}$. Note that throughout this paper, the $r_{\rm BLR,i}$ is fixed at $5000~R_{\rm g}$, while $r_{\rm BLR,o}$ is fixed at $10^{5}~R_{\rm g}$ ($R_{\rm g}$ is the Gravitational radius, $R_{\rm g}={2GM}/{c^{2}}$). This constrain is based on the assumptions of canonical blazars as follows: (1) the accretion disk is a standard disk \citep{1973A&A....24..337S}. The inner and outer radius are $R_{\rm disk,i}=6~R_{\rm g}$ and $R_{\rm disk,o}=1000~R_{\rm g}$, respectively \citep{2009MNRAS.397..985G}; (2) the accretion disk is within the cavity of the BLR, while the $r_{\rm BLR,o}$ is fixed at $10^{5}~R_{\rm g}$ \citep{2009ApJ...692...32D}. The blackbody temperature of the inner region of the accretion disk is assumed to be $T=10^{5}$\,K. The detailed mathematical treatments on the BLR are presented in Appendix.

The accretion disk emits a total luminosity of $L_{\rm d}$ uniformly toward all directions, the BLR clouds intercept a fraction of illuminating continuum and reprocess them to the emission lines and the diffuse continuum. Moreover, as stated by \citet{2010ApJ...717L.118P}, the BLR also emits the recombination continua of hydrogen and He II, they could cause jumps in the $\gamma$-ray opacity at $\sim 19.2$ and 4.8\,GeV. However, in this work, we do not consider the contributions from the recombination continua to the $\gamma$-ray absorption. For simplicity, we assume that the fractions $f_{\rm cov}$ and $\tau_{\rm BLR}$ of $L_{\rm d}$ are reprocessed into the emission lines $j_{\rm line}$ and the diffuse continuum $j_{\rm cont}$ respectively, $f_{\rm cov}$ and $\tau_{\rm BLR}$ are set to be 0.05, i.e., the total reprocessing fraction is fixed at 0.1. In order to evaluate the total emission line flux from the BLR, we consider the 35 components of emission lines. The sum of line ratios is given by $N_{\nu}=555.77$ relative to $\rm{Ly}\alpha$ line ratio ($N_{\nu,\rm Ly\alpha}=100$). Therefore the flux contributed by a certain emission line equals
\begin{eqnarray}
F_{\nu}&=&\frac{N_{\nu}}{555.77}F_{\rm BLR},
\end{eqnarray}
where $F_{\rm BLR}$ is the total emission line flux contributing to $\gamma$-ray emission region. Throughout this paper, the luminosity of the central disk is fixed at $L_{\rm d}=5\times 10^{46}$~erg~$s^{-1}$, corresponding roughly to the intermediate value of the luminosity of FSRQs.

\subsection{The $\gamma$-ray optical depth}
According to the assumptions used by \citet{2006ApJ...653.1089L}, the luminosity emitted by accretion disk are reprocessed into the emission lines and the diffuse continuum.
The emissivity of the emission lines is
\begin{eqnarray}
j_{\rm line}(r)&=&\frac{\emph{f}_{\rm cov}~L_{\rm d}~r^{2q-p-2}}{16\pi^{2}\int_{r_{\rm BLR,i}}^{r_{\rm BLR,o}}r^{2q-p}dr}.
\end{eqnarray}
The emissivity of the diffuse continuum is
\begin{eqnarray}
j_{\rm cont}(r)&=&\frac{L_{\rm d}~\tau_{\rm BLR}~r^{-s-2}}{16\pi^{2}\int_{r_{\rm BLR,i}}^{r_{\rm BLR,o}}r^{-s}dr},
\end{eqnarray}
where the number densities of the clouds and electrons are assumed to follow the power-law distributions, e.g., $n_{\rm c}=n_{\rm c0}(r/r_{\rm BLR,i})^{-p}$ and $n_{\rm e}=n_{\rm e0}(r/r_{\rm BLR,i})^{-s}$, where $n_{\rm c0}$ and $n_{\rm e0}$ are the number densities of clouds and electrons at $r_{\rm BLR,i}$, respectively. For the radial distribution of clouds, the same assumption is
adopted, such as $R_{\rm cd}=R_{\rm cd0}(r/r_{\rm BLR,i})^{q}$, where $R_{\rm cd0}$ is the radius of clouds at $r_{\rm BLR,i}$. In present paper, we adopt the power-law exponents preferred by \citet{1999ApJ...524...71K} to be $p=1.5, s=1$ and $q=1/3$.

The intensity of radiation with the angle $\theta$ respect to the jet axis at position $R$ is
\begin{eqnarray}
I(R,\theta)&=&\int_{l_{\rm min}}^{l_{\rm max}} j(r)dl,
\end{eqnarray}
where $r^{2}=R^{2}-2Rl\cos\theta+l^{2}$, $j(r)$ is the total emissivity that has $j(r)=j_{\rm line}(r)+j_{\rm cont}(r)$ and vanishes when $r>r_{\rm BLR,o}$ or $r<r_{\rm BLR,i}$.

The soft photon energy density of the BLR is given by

\begin{eqnarray}
U(R)&=&\frac{2\pi}{c}\int_{\theta_{\rm min}}^{\theta_{\rm max}} I(R,\theta) \sin\theta d\theta.
\end{eqnarray}

The photon number densities associated with the emission lines and the diffuse continuum are
\begin{eqnarray}
n_{\rm line}(R,\nu,\Omega)&=&\frac{I_{\rm line}(R,\nu,\theta)}{h\nu c},
\nonumber\\
n_{\rm cont}(R,\nu,\Omega)&=&\frac{2~\nu^{2}}{c^{3}[exp(h\nu/kT)-1]}n(R,\theta),
\end{eqnarray}
where $I_{\rm line}(R,\nu,\theta)$ is the monochromatic intensity, and $n(R,\theta)$ is the normalization factor.

At last, the $\gamma$-ray optical depth is calculated by
\begin{eqnarray}
\tau_{\gamma\gamma}(\epsilon_{\gamma})&=&\int_{R_{\rm o}}^{R_{\rm max}}dR\int_{\epsilon_{\rm l}}^{\epsilon_{\rm u}} d\epsilon \int_{\theta_{\rm min}}^{\theta_{\rm max}}d\theta
\nonumber\\
&\times&\sigma(\epsilon_{\gamma},\epsilon,\theta)n(R,\epsilon,\Omega)(1-\cos\theta),
\end{eqnarray}
where $\epsilon$ is the dimensionless energy of soft photon, $n(R,\epsilon,\theta)$ is the soft photon number density at position $R$ with the angle $\theta$ of integral path respect to jet axis. In the calculation of $\gamma-\gamma$ attenuation optical depth, the soft photon frequencies are from $\nu_{\rm l}=10^{14.6}$\,Hz to $\nu_{\rm u}=10^{16.5}$\,Hz. The upper limit over $R$ is set to be 2\,pc in all the cases. $\sigma(\epsilon_{\gamma},\epsilon,\theta)$ is the pair-creation cross section, and the threshold of $\sigma(\epsilon_{\gamma},\epsilon,\theta)$ is determined by
\begin{eqnarray}
\epsilon_{\gamma} &=& \frac{2}{(1-\cos\theta)\epsilon}.
\end{eqnarray}

In order to increase the threshold of $\gamma$-ray photons in collisions, we can reduce the factor $(1-\cos\theta)$. In a ``flat" geometry of the BLR, the decrease of $\alpha$ will reduce the probability of ``head-on" collisions and increase the threshold of the $\gamma$-ray photons.

\section{Results}
\begin{figure}
  \centerline{
    \FigureFile(150mm,80mm){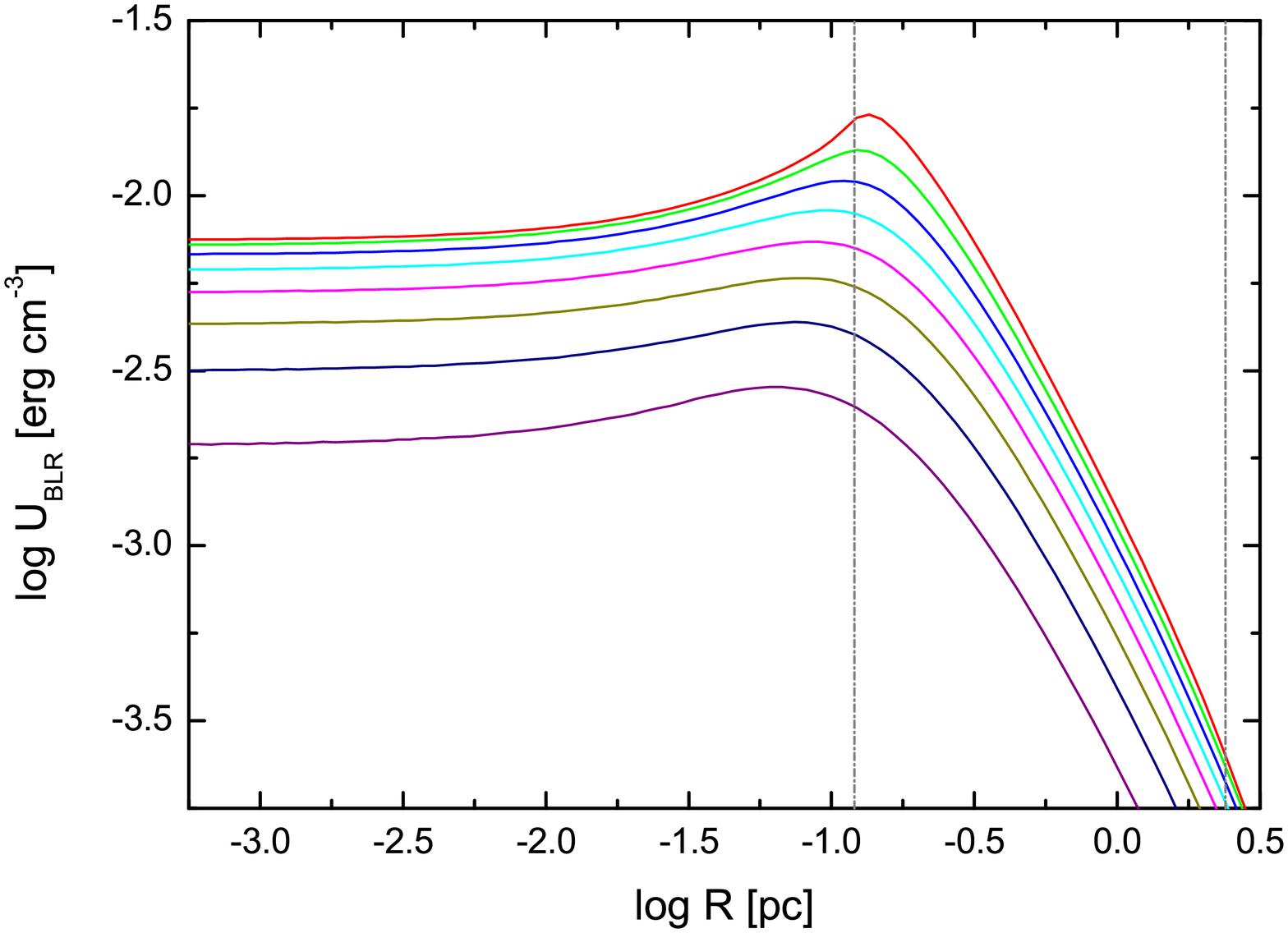}}
  \caption{Dependence of the energy density $U_{\rm BLR}$ on the radial distance $R$ from the central engine. The different colored lines correspond to the different $\alpha$, from top to down, which are $85^{\circ}, 75^{\circ}$, $65^{\circ}$, $55^{\circ}$, $45^{\circ}$, $35^{\circ}$, $25^{\circ}$, $15^{\circ}$, respectively. The two vertical lines correspond to the inner and outer radius of the BLR.}
  \label{fig:U_BLR}
\end{figure}

To reduce the number of parameters, we keep the mass of central supermassive black hole\,(BH) to be $M_{\rm BH}=5\times 10^{8} M_{\rm \odot}$, which is a typical mass of BH in AGNs. The total energy density distributions of the BLR are shown in figure \ref{fig:U_BLR}. It is obvious that the aperture angle $\alpha$ has dramatic impact on the energy density distributions. Actually, with the increase of $\alpha$, the illuminating continuum intercepted by the BLR clouds increases, leading to the increase of the flux of the reprocessed emissions and the diffuse continuum. The energy density is almost constant within the cavity of the BLR, but at distance nearly close to $r_{\rm BLR,i}$, the energy density increases because the anisotropic effect of the BLR geometry becomes more apparent. Furthermore, the energy density declines outside $r_{\rm BLR,i}$, but when $\alpha$ is relatively low, the energy density declines even if the distance is less than $r_{\rm \rm BLR,i}$. In the case of small $\alpha$, the BLR is disc-like, the energy density is inversely proportional to the distance. In contrast, in larger $\alpha$ nearly equaling to $90^{\circ}$, the energy density is contributed by clouds within the BLR and will decrease along the radial distance due to the obscure effect of the BLR clouds. Overall, the location of the peaks of the energy density shifts toward larger distance with the increase of $\alpha$.

\begin{figure}
  \centerline{
    \FigureFile(150mm,80mm){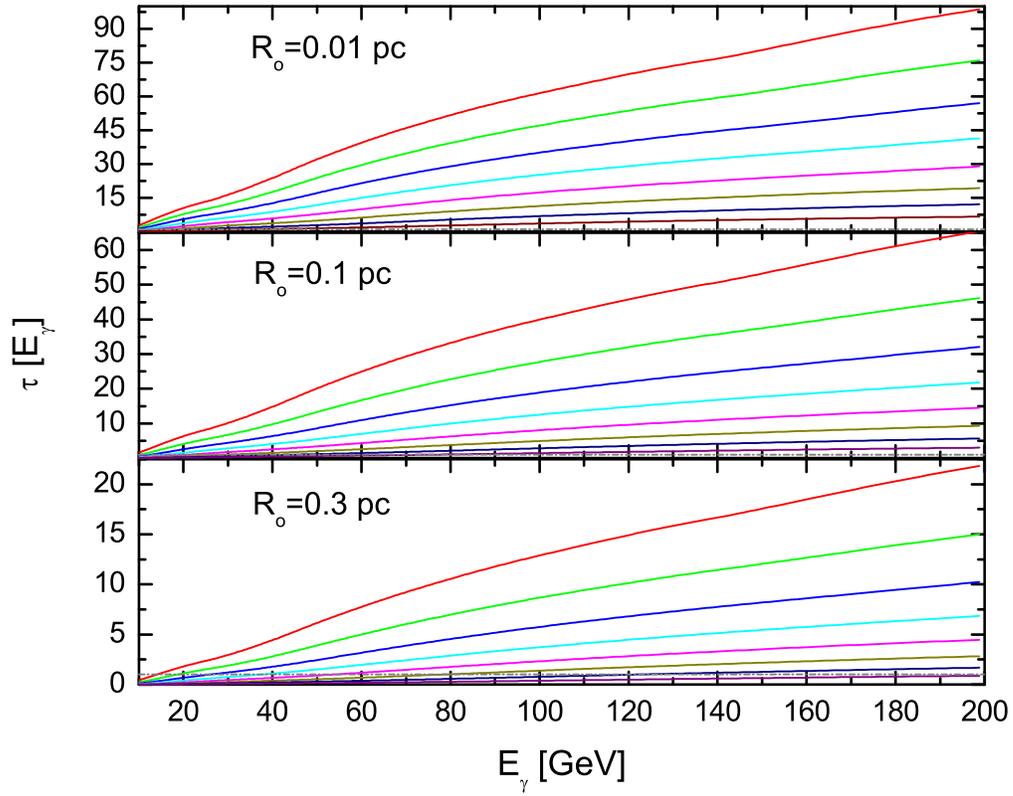}}
  \caption{Dependence of the $\gamma$-ray optical depth $\tau$ on $\gamma$-ray photon energy $E_{\gamma}$. The different colored lines correspond to different $\alpha$, from top to down, which are $85^{\circ}, 75^{\circ}$, $65^{\circ}$, $55^{\circ}$, $45^{\circ}$, $35^{\circ}$, $25^{\circ}$, $15^{\circ}$, respectively. The horizonal line corresponds to $\tau=1$.}
  \label{fig:tau_E}
\end{figure}

The soft photons from the BLR are not only scattered by relativistic electrons in the jet to produce high-energy component in the SEDs of FSRQs, are but also the absorbing sources of $\gamma$-rays via the processes of pair-creation. \citet{2006ApJ...653.1089L} found that the $\gamma$-ray photons with energies from 10 to 200\,GeV cannot escape from the diffuse field of the BLR with a spherical shell geometry. In this paper, we extend a ``closed" geometry to a ``flat" geometry for calculating the $\gamma$-rays absorption. Because the emissivity of diffuse and line emissions is proportional to the disk luminosity $L_{\rm d}$, the optical depth $\tau_{\gamma\gamma}$ is also proportional to $L_{\rm d}$. In Figure \ref{fig:tau_E}, we plot the optical depths versus $\gamma$-ray energies. The $\gamma$-ray absorption are shown at three different distances, e.g., $R_{\rm o}=0.01,~0.1,~0.3$\,pc. It is shown that the aperture angle $\alpha$ affects the $\gamma$-rays optical depth $\tau$ greatly. From the top panel in Figure \ref{fig:tau_E}. The emitting region locates at $R_{\rm o}$=0.01\,pc, we can see that even if $\alpha$ is as low as $15^{\circ}$, the optical depth of the $\gamma$-ray photons with energies $\sim35$\,GeV approaches unity, the photons with energies lower than this critical energy can escape out the diffuse field. From middle panel in Figure \ref{fig:tau_E}, when $R_{\rm o}$ approaches $r_{\rm BLR,i}$ and $\alpha$ equals to $15^{\circ}$, the $\gamma$-ray photons with energies lower than critical energy $\sim75$\,GeV can escape out the diffuse field. Increasing $\alpha$ to $25^{\circ}$, the critical energy is down to $\sim50$\,GeV. Accordingly, $\alpha$ grows up to $35^{\circ}$, the critical energy is down to $\sim30$\,GeV. When $R_{\rm o}$=0.3\,pc within the BLR clouds and $\alpha$ is less than $35^{\circ}$, the $\gamma$-ray photons with energies up to $\sim85$\,GeV can escape out the diffuse field.

\begin{figure}
  \centerline{
    \FigureFile(150mm,80mm){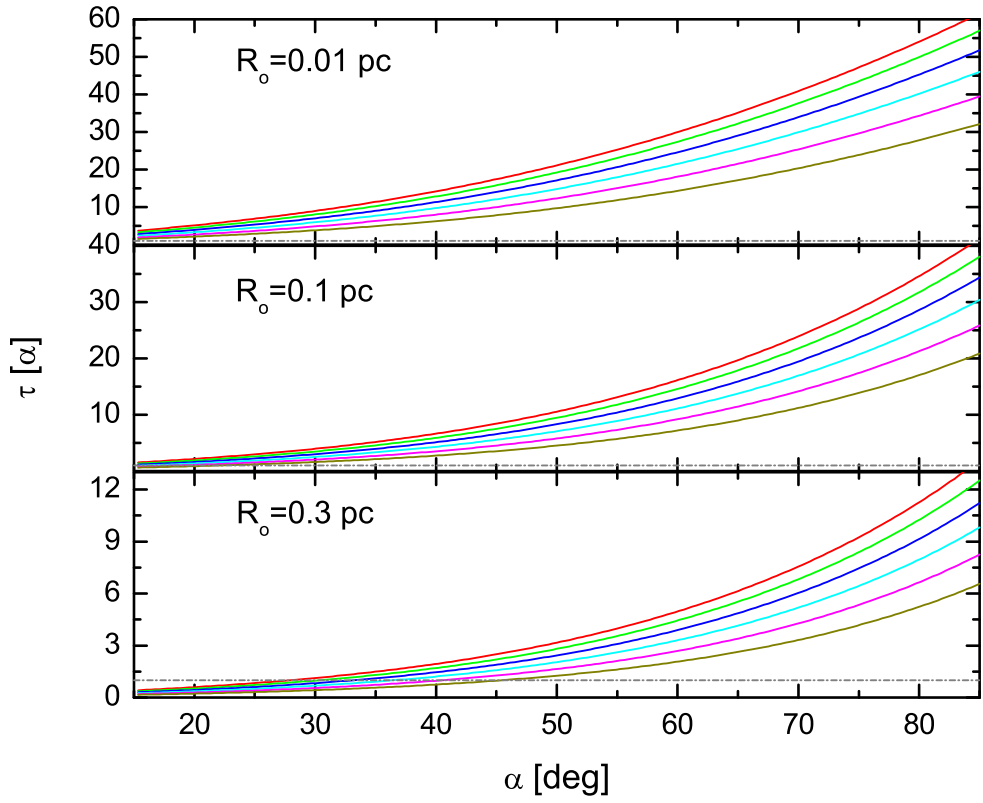}}
  \caption{Dependence of the $\gamma$-ray optical depth $\tau$ on the aperture angle $\alpha$. The different colored lines correspond to different photon energies, from top to down, $E_{\gamma}$=100, 90, 80, 70, 60, 50\,GeV, respectively. The horizonal line corresponds to $\tau=1$.}
  \label{fig:tau_alpha}
\end{figure}

We calculate the optical depths of $\gamma$-ray photons in the framework of ``flat" geometry of the BLR, in which ``flatness" is characterized by the aperture angle $\alpha$. For illustrating the dependence of $\tau$ on $\alpha$, we consider three positions as above, $R_{\rm o}=0.01,~0.1, ~0.3$\,pc, while the energies of $\gamma$-ray photons are considered from 50 to 100\,GeV energy bands, the results are shown in Figure \ref{fig:tau_alpha}.
It shows that when the $\gamma$-ray emitting region locates at position $R_{\rm o}=0.01$\,pc, the energy density of the diffuse field is large and opaque to all of the $\gamma$-ray photons with energies beyond 50\,GeV as shown in the upper panel in figure \ref{fig:tau_alpha}. In middle panel, the $\gamma$-ray emitting region is close to the inner boundary of the BLR, the $\gamma$-ray photons undergo the short path before they escape out the BLR under a certain $\alpha$. We can observe the $\gamma$-ray photons with energy 50\,GeV in the case of $\alpha \sim 24^{\circ}$, 60\,GeV in the case of $\alpha \sim 20^{\circ}$, 70\,GeV in the case of $\alpha \sim 17^{\circ}$. When $\gamma$-ray emitting region locates within the clouds of the BLR, as long as the aperture angle $\alpha\lesssim30^{\circ}$, the $\gamma$-ray photons from 50\,GeV to 100\,GeV will be observed.

Because the optical depth depends on the height of the emitting region and the aperture angle, when $\tau_{\gamma\gamma}=1$, we can find a critical height $R_{\rm c}$ of $\gamma$-ray transparency for the aperture angle $\alpha$ shown in Figure \ref{fig:Rc_alpha}, for $\gamma$-ray photons with the energies from 50\,GeV to 100\, GeV. It is shown that the critical aperture angle $\alpha_{\rm c}$ moves toward smaller angle with the increase of the $\gamma$-rays energies, because the decrease of $\alpha$ will reduce the probability of ``head-on" collisions, leading the increase of threshold energy. For the $\gamma$-ray with 100\,GeV,  $R_{\rm c}=0.1$\,pc, the critical aperture angle $\alpha_{\rm c}$ is nearly $15^{\circ}$, and the geometry of the BLR becomes disc-like.

\begin{figure}
  \centerline{
    \FigureFile(150mm,80mm){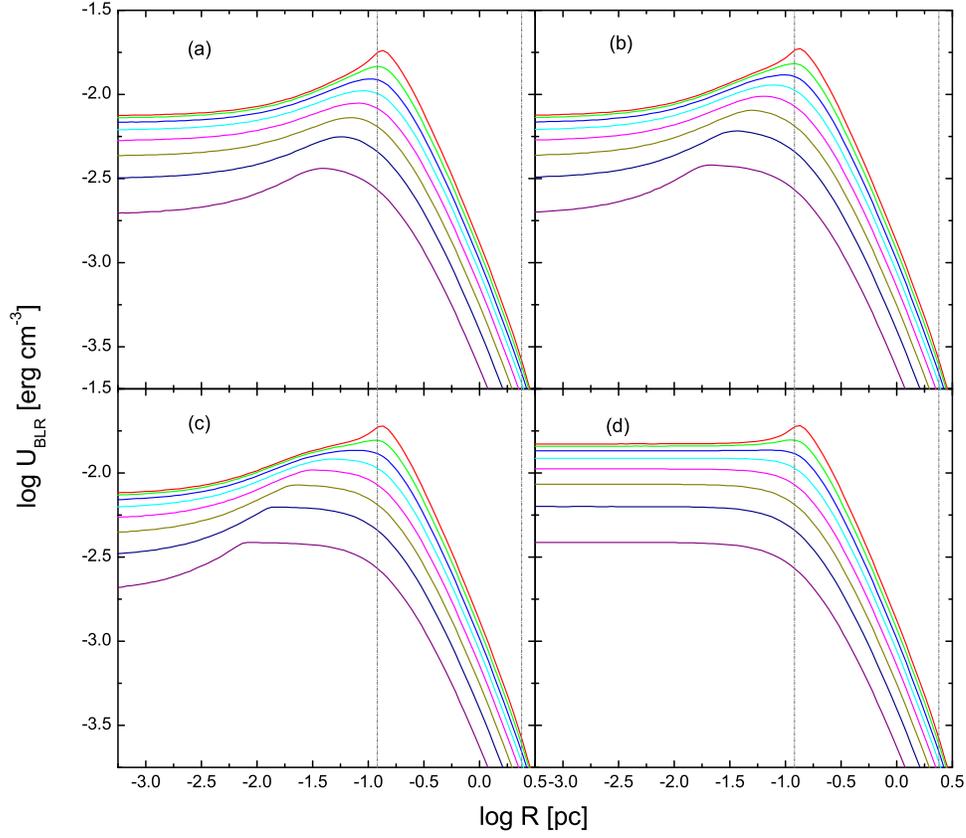}}
  \caption{Dependence of the energy density of the diffuse field $U_{\rm BLR}$ on the radial distance $R$ from the central engine. The different colored lines correspond to the different $\alpha$, from top to down, which are $85^{\circ}, 75^{\circ}$, $65^{\circ}$, $55^{\circ}$, $45^{\circ}$, $35^{\circ}$, $25^{\circ}$, $15^{\circ}$. We change the outer radius of the accretion disk in the panels (a)-(d) as $R_{\rm disk,o}/r_{\rm BLR,i}=0.6,~0.4,~0.2,~0$, respectively. The two vertical dashed lines represents the inner and outer radius of the BLR.}
  \label{fig:U_BLR_ratio}
\end{figure}

\begin{figure}
  \centerline{
    \FigureFile(150mm,80mm){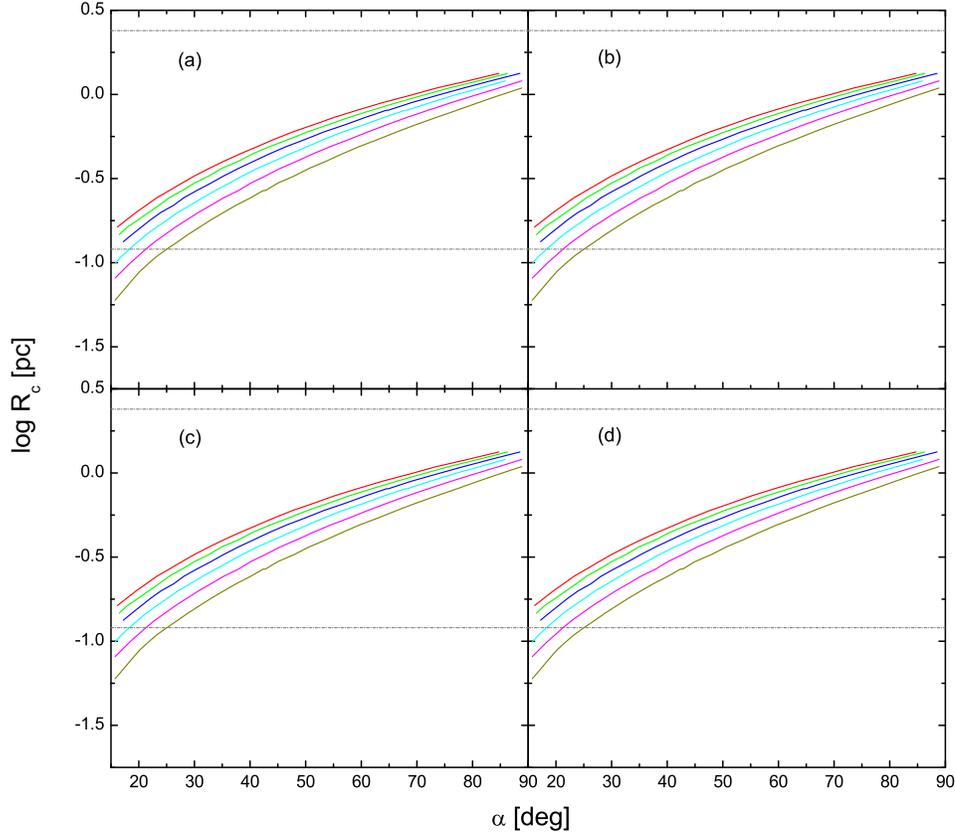}}
  \caption{Setting $\tau=1$, dependence of critical position $R_{\rm c}$ on the aperture angle $\alpha$ according with figure \ref{fig:U_BLR_ratio}. The different colored lines correspond to different $\gamma$-ray photon energies, from top to down, $E_{\gamma}$=100, 90, 80, 70, 60, 50\,GeV, respectively. We change the outer radius of the accretion disk in the panels (a)-(d) as $R_{\rm disk,o}/r_{\rm BLR,i}=0.6,~0.4,~0.2,~0$, respectively. The two horizonal dashed lines represents the inner and outer radius of the BLR.}
  \label{fig:Rc_ratio}
\end{figure}

The above results are based on the assumption that the outer radius $R_{\rm disk,o}$ of accretion disk extends to the inner radius $r_{\rm BLR,i}$ of the BLR, e.g., $R_{\rm disk,o}=r_{\rm BLR,i}$. In the following, we try to consider the effect of the disk size on soft photon fields and $\gamma$-ray absorption. The results are shown in Figure \ref{fig:U_BLR_ratio} for $U_{\rm BLR}$ versus $R$, in which $\alpha$ is set to $15^{\circ}$, and in Figure \ref{fig:Rc_ratio} for $R_{\rm c}$ versus $\alpha$, in which the outer radius of disk varys as $R_{\rm disk,o}/r_{\rm BLR,i}=0.6,~0.4,~0.2,~0$. The disk size mainly influences the energy density of the diffuse photons field inside the BLR. When $R_{\rm disk,o}/r_{\rm BLR,i}=0$, e.g., the central engine is a point source of radiation that isotropically emits and radially locates behind the jet, and the energy density inside the BLR becomes a constant. It is noted that the disk size does not affect the relation of $R_{\rm c}$ and $\alpha$. It is evident that BLR clouds do not exist near the jet axis under the disk when $\alpha$ is small, and $R_{\rm disk,o}$ does not change the energy density of the diffuse photons field. Furthermore, when $\alpha$ is large, the critical position $R_{\rm c}$ will be within the BLR clouds, and the energy density keeps constant due to the obscure effect of BLR clouds.

\section{Discussion and conclusions}
\begin{figure}
  \centerline{
    \FigureFile(150mm,80mm){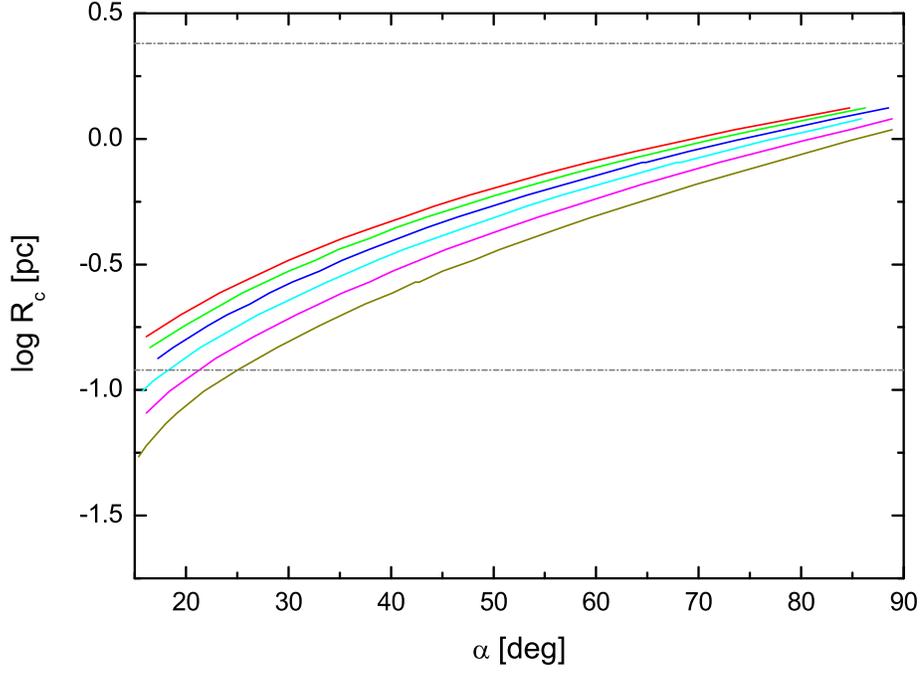}}
  \caption{Setting $\tau=1$, dependence of critical position $R_{\rm c}$ on the aperture angle $\alpha$. The different colored lines correspond to different $\gamma$-ray photon energies, from top to down, $E_{\gamma}$=100, 90, 80, 70, 60, 50\,GeV, respectively. The two horizonal dashed lines represents the inner and outer radius of the BLR.}
  \label{fig:Rc_alpha}
\end{figure}

The geometry of the BLR is usually assumed to be an isotropic shell, the absorption of the $\gamma$-ray photons by the diffuse field is serious when the $\gamma$-ray emitting region locates within the cavity of the BLR \citep{2006ApJ...653.1089L,2012arXiv1209.2291T}. The present work extends an isotropic BLR to a ``flat" BLR described by a aperture angle $\alpha$. In the previous work \citep{2012arXiv1209.2291T}, the BLR clouds are assumed to intercept a fraction $\rm{C}=\Omega_{\rm BLR}/2\pi=0.1$ of the illuminating continuum, where $\Omega_{\rm BLR}$ is the solid angle surrounded by the BLR clouds, e.g., $\Omega_{\rm BLR}=2\pi \sin\alpha_{\rm min}$. The minimum aperture angle is then given by $\alpha_{\rm min} = \arcsin {\rm C}$, we get $\alpha_{\rm min} \simeq 5.7^{\circ}$. Giving this limit, we can determine the upper limit $\alpha_{\rm max}$ that increases with the decline of $\gamma$-ray energy shown in Figure \ref{fig:Rc_alpha}. Alternatively, the geometry of the BLR could be reflected by the deprojection factor \emph{f}. In the commonly assumed isotropic shell, $<\emph{f}>=\sqrt3/2$. When the geometry is disc-like \citep{2001MNRAS.327..199M}
\begin{eqnarray}
\emph{f}=0.5\big[(\frac{H}{R})^{2}+sin^{2}\theta\big]^{-1/2},
\end{eqnarray}
where $H/R$ is the scale ratio of the BLR, $H$ is the thickness, $R$ is the radius, and $\theta$ is the angle between the normal to the disk and the line of sight. We then find that the relationship between $\alpha$ and $\emph{f}$ is given by
\begin{eqnarray}
\tan\alpha=\big[\frac{1}{16\emph{f}~^{2}}-\frac{1}{4}\sin^{2}\theta\big]^{1/2}.
\end{eqnarray}
When we take $f=2.0$ according to the result given by \citet{2011MNRAS.413...39D}, and $\theta \simeq 1/\delta_{\rm D}$ for blazars, in which $\delta_{\rm D}$ is the Doppler factor from 10 to 50 randomly, we obtain $\alpha\approx7^{\circ}$, indicating that the geometry of the BLR is rather flat. Under this case, almost all the $\gamma$-ray photons can escape out the diffuse photons field.

In summary, we extend the spherical geometry of the BLR to the flat geometry characterized by the aperture angle $\alpha$. In calculations, we take the central BH mass of $5\times 10^{8}~M_{\rm \odot}$, the inner and the outer radius of the BLR to be $5\times10^{3}~R_{\rm g}$ and $10^{5}~R_{\rm g}$, and the luminosity of the central disk to be $L_{\rm d}=5\times 10^{46}$~erg~s$^{-1}$. The diffuse photon energy density decreases with the decline of $\alpha$, but it keeps constant inside $r_{\rm BLR,i}$ and declines outside $r_{\rm BLR,i}$ for a given $\alpha$. The optical depth $\tau$ of $\gamma$-rays not only depends on $R_{\rm o}$ and $\gamma$-ray energy, but also on $\alpha$ strongly. Setting $\tau=1$, we can obtain the relation between the critical distance $R_{\rm c}$ and the aperture angle $\alpha$ for different energies. We find that when the geometry of the BLR is flat, the $\gamma$-rays with specified energies can escape away the cavity of the BLR.

We acknowledge the financial supports from the National Basic Research Program of China (973 Program 2009CB824800), the National Natural Science Foundation of China 11133006, 11163006, 11173054, and the Policy Research Program of Chinese Academy of Sciences (KJCX2-YW-T24).

\begin{figure}
  \centerline{
    \FigureFile(100mm,80mm){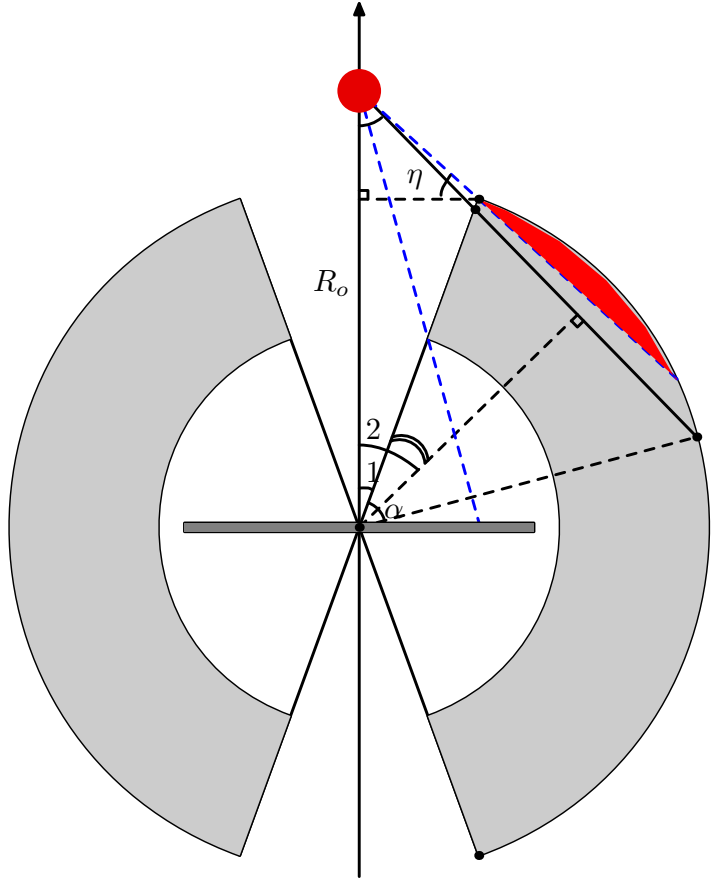}}
  \caption{The geometry and mathematical treatments of the BLR clouds. The central strip illustrates the accretion disk, and the outer shaded
  symmetrical structures show the BLR clouds characterized by the aperture angle $\alpha$. The red blob represents the $\gamma$-ray
  emission region, located at $R_{o}$ at a certain time.
  The region surrounded by two dashed blue lines is integrated to calculate the contributions of the diffuse field to $\gamma$-rays absorption, while the red region is treated separately when $\alpha$ is very large. $\eta_{\rm c}$ is the critical position whether the red region is integrated or not, see the text in detail.
  }
\label{fig:BLR}
\end{figure}

\begin{center}
{Appendix} \\
{Mathematical treatments on the BLR}
\end{center}
As shown in Figure \ref{fig:BLR}, we firstly perform integration for the shaded region surrounded by two blue dashed lines, the black solid line is the integral path to $\emph{l}$~ that has a angle of $\theta$ with respect to the jet axis, and $l_{\rm min}$, $l_{\rm max}$ are the lower and upper limits, respectively. The integral limits are given by

\begin{eqnarray}
\theta_{\rm min}&=&0,
\nonumber\\
\theta_{\rm max}&=&\arccos\big[\frac{R^{2}+x_{\rm l}^{2}-r_{\rm BLR,o}^{2}}{2Rx_{\rm l}}\big],
\nonumber\\
l_{\rm min}&=&R\cos\theta+R\sin \theta\tan \theta_{\rm \delta},
\end{eqnarray}
and
\begin{eqnarray}
\theta_{\rm cr}^{\rm min}&=&\arccos\big[\frac{R}{\sqrt{R^{2}+r_{\rm BLR,i}^{2}}}\big],
\nonumber\\
\theta_{\rm cr}^{\rm apex}&=&\arccos\big[\frac{R^{2}+\xi^{2}-r_{\rm BLR,o}^{2}}{2R\xi}\big].
\end{eqnarray}

Here $\theta_{\rm cr}^{\rm min}$ is the angle that jet axis makes with the line linking the emitting region and outer boundary of the disk, while $\theta_{\rm cr}^{\rm apex}$ is the angle that jet axis makes with the line linking the emitting region and apex ``N". $l_{\rm max}$ is determined by \\
if $\theta \geq \theta_{\rm cr}^{\rm apex} \geq \theta_{\rm cr}^{\rm min}$:
\begin{eqnarray}\label{eq:l_begin}
l_{\rm max}&=& R\cos\theta+R\big[\cos^{2}\theta+(\frac{r_{\rm BLR,o}}{R})^{2}-1\big]^{1/2},
\end{eqnarray}
if $\theta_{\rm cr}^{\rm apex} \geq \theta \geq \theta_{\rm cr}^{\rm min}$:
\begin{eqnarray}
l_{\rm max}&=& R\cos\theta+R\sin\theta\tan(\alpha+\theta),
\end{eqnarray}
if $\theta_{\rm cr}^{\rm apex} > \theta_{\rm cr}^{\rm min} > \theta$:
\begin{eqnarray}
l_{\rm max}&=& R/\cos\theta,
\end{eqnarray}
if $\theta \geq \theta_{\rm cr}^{\rm min} \geq \theta_{\rm cr}^{\rm apex}$:
\begin{eqnarray}
l_{\rm max}&=& R\cos\theta+R\big[\cos^{2}\theta+(\frac{r_{\rm BLR,o}}{R})^{2}-1\big]^{1/2},
\end{eqnarray}
if $\theta_{\rm cr}^{\rm min} > \theta_{\rm cr}^{\rm apex} > \theta$:
\begin{eqnarray}\label{eq:l_end}
l_{\rm max}&=& R/\cos\theta,
\end{eqnarray}
where
\begin{eqnarray}
x_{\rm l}&=&\big[(r_{\rm BLR,o}\cos\alpha)^{2}+(R-r_{\rm BLR,o}\sin\alpha)^{2}\big]^{1/2},
\nonumber\\
\theta_{\rm \delta}&=&\angle_{\rm 1}-\angle_{2},
\nonumber\\
\xi&=&\big[R^{2}+r_{\rm BLR,o}^{2}-2Rr_{\rm BLR,o}\cos(\pi/2+\alpha)\big]^{1/2}.
\end{eqnarray}

It is noted that when we integrate the shaded region surrounded by two blue dashed lines, the red region will not be integrated, because with the increase of $\alpha$, the line linking $\gamma$-ray emitting region and apex ``M" becomes more steeper, leading to the loss of the red region. For compensation to above calculation, we consider a critical position that satisfies the condition: $\alpha+\eta_{\rm c}=\pi/2$, if $\alpha+\eta_{\rm c}\leq \pi/2$, $R\leq r_{\rm BLR,o}\sin{\alpha}$, the contributions from red region to the soft seed photon field could be neglected. Inversely, we have to integrate the red region. The integral limits of the red region are given by
\begin{eqnarray}
\theta_{\rm min}&=&\arccos\big[\frac{R^{2}+x_{\rm l}^{2}-r_{\rm BLR,o}^{2}}{2Rx_{\rm l}}\big],
\nonumber\\
\theta_{\rm max}&=&\arccos\big[\frac{\sqrt{R^{2}-r_{\rm BLR,o}^{2}}}{R}\big],
\end{eqnarray}
\begin{eqnarray}
l_{\rm min}&=&R\cos\theta-R\big[\cos^{2}\theta+(\frac{r_{\rm BLR,o}}{R})^{2}-1\big]^{1/2},
\end{eqnarray}
while $l_{\rm max}$ is taken as Equation (\ref{eq:l_begin})--(\ref{eq:l_end}).


\newpage

\begin{thebibliography}{}
\bibitem[Tavecchio
\& Ghisellini(2012)]{2012arXiv1209.2291T} Tavecchio, F., \& Ghisellini, G.\ 2012, arXiv:1209.2291


\bibitem[Stern
\& Poutanen(2011)]{2011MNRAS.417L..11S} Stern, B.~E., \& Poutanen, J.\ 2011, \mnras, 417, L11


\bibitem[Ghisellini et al.(2011)]{2011MNRAS.414.2674G} Ghisellini, G.,
Tavecchio, F., Foschini, L., \& Ghirlanda, G.\ 2011, \mnras, 414, 2674


\bibitem[Decarli et al.(2011)]{2011MNRAS.413...39D} Decarli, R., Dotti, M.,
\& Treves, A.\ 2011, \mnras, 413, 39


\bibitem[Poutanen
\& Stern(2010)]{2010ApJ...717L.118P} Poutanen, J., \& Stern, B.\ 2010, \apjl, 717, L118


\bibitem[Abdo et al.(2010)]{2010ApJ...716...30A} Abdo, A.~A., Ackermann,
M., Agudo, I., et al.\ 2010, \apj, 716, 30


\bibitem[Ghisellini
\& Tavecchio(2009)]{2009MNRAS.397..985G} Ghisellini, G., \& Tavecchio, F.\ 2009, \mnras, 397, 985


\bibitem[Bentz et al.(2009)]{2009ApJ...697..160B} Bentz, M.~C., Peterson,
B.~M., Netzer, H., Pogge, R.~W., \& Vestergaard, M.\ 2009, \apj, 697, 160


\bibitem[Dermer et al.(2009)]{2009ApJ...692...32D} Dermer, C.~D., Finke,
J.~D., Krug, H., \& B$\ddot{o}$ttcher, M.\ 2009, \apj, 692, 32


\bibitem[Tavecchio
\& Mazin(2009)]{2009MNRAS.392L..40T} Tavecchio, F., \& Mazin, D.\ 2009, \mnras, 392, L40


\bibitem[Ghisellini
\& Tavecchio(2008)]{2008MNRAS.387.1669G} Ghisellini, G., \& Tavecchio, F.\ 2008, \mnras, 387, 1669


\bibitem[Decarli et al.(2008)]{2008MNRAS.386L..15D} Decarli, R., Dotti, M.,
Fontana, M., \& Haardt, F.\ 2008, \mnras, 386, L15


\bibitem[Tavecchio
\& Ghisellini(2008)]{2008MNRAS.386..945T} Tavecchio, F., \& Ghisellini, G.\ 2008, \mnras, 386, 945


\bibitem[Kaspi et al.(2007)]{2007ApJ...659..997K} Kaspi, S., Brandt, W.~N.,
Maoz, D., et al.\ 2007, \apj, 659, 997


\bibitem[Celotti et al.(2007)]{2007MNRAS.375..417C} Celotti, A.,
Ghisellini, G., \& Fabian, A.~C.\ 2007, \mnras, 375, 417


\bibitem[Labita et al.(2006)]{2006MNRAS.373..551L} Labita, M., Treves, A.,
Falomo, R., \& Uslenghi, M.\ 2006, \mnras, 373, 551


\bibitem[Liu
\& Bai(2006)]{2006ApJ...653.1089L} Liu, H.~T., \& Bai, J.~M.\ 2006, \apj, 653, 1089


\bibitem[Kaspi et al.(2005)]{2005ApJ...629...61K} Kaspi, S., Maoz, D.,
Netzer, H., et al.\ 2005, \apj, 629, 61


\bibitem[Grupe
\& Mathur(2004)]{2004ApJ...606L..41G} Grupe, D., \& Mathur, S.\ 2004, \apjl, 606, L41


\bibitem[Donea
\& Protheroe(2003)]{2003APh....18..377D} Donea, A.-C., \& Protheroe, R.~J.\ 2003, Astroparticle Physics, 18, 377


\bibitem[McLure
\& Dunlop(2001)]{2001MNRAS.327..199M} McLure, R.~J., \& Dunlop, J.~S.\ 2001, \mnras, 327, 199


\bibitem[Hayashida(2000)]{2000NewAR..44..419H} Hayashida, K.\ 2000, \nat,
44, 419


\bibitem[Kaspi et al.(2000)]{2000ApJ...533..631K} Kaspi, S., Smith, P.~S.,
Netzer, H., et al.\ 2000, \apj, 533, 631


\bibitem[Kaspi
\& Netzer(1999)]{1999ApJ...524...71K} Kaspi, S., \& Netzer, H.\ 1999, \apj, 524, 71


\bibitem[Urry
\& Padovani(1995)]{1995PASP..107..803U} Urry, C.~M., \& Padovani, P.\ 1995, \pasp, 107, 803


\bibitem[Blandford
\& McKee(1982)]{1982ApJ...255..419B} Blandford, R.~D., \& McKee, C.~F.\ 1982, \apj, 255, 419


\bibitem[Shakura
\& Sunyaev(1973)]{1973A&A....24..337S} Shakura, N.~I., \& Sunyaev, R.~A.\ 1973, \aap, 24, 337


\end{thebibliography}
\end{document}